\documentclass[prb]{revtex4}

\usepackage{epsf}

\newcommand{\be}{\begin{equation}}
\newcommand{\ee}{\end{equation}}
\newcommand{\bea}{\begin{eqnarray}}
\newcommand{\eea}{\end{eqnarray}}

\newcommand{\p}{\partial}
\newcommand{\re}{{\rm e}}
\newcommand{\rd}{{\rm d}}

\begin{document}
\title{ On the quasiparticle description of c=1 CFTs}
\author{Davide Controzzi 
\footnote{ On leave from  {\sl International School for 
Advanced Studies, Trieste. }} and Kareljan Schoutens}
\affiliation
{Institute for Theoretical Physics, 
Valckenierstraat 65, 1018 XE Amsterdam,
The Netherlands}

\begin{abstract}
\par
%We show that the description of $c=1$ Conformal Field Theory in terms 
%of quasiparticles satisfying fractional statistics can be obtained 
%from the sine-Gordon model with a chemical potential $A$, in the limit where
%$A \gg M$. We provide a direct calculation of the 2-particle S-matrix 
%using Korepin's method. 
%The CFT quasiparticles are related to the excitations 
%of the Calogero-Sutherland (CS) model. We reconsider the computation of 
%the CS S-matrix in terms of particles with fractional charge. 
We show that the description of $c=1$ Conformal Field Theory in terms 
of quasiparticles satisfying fractional statistics can be obtained 
from the sine-Gordon model with a chemical potential $A$, in the limit where
$A \gg M$. These quasiparticles are related to the excitations 
of the Calogero-Sutherland (CS) model. We  
provide a direct calculation of their 2-particle S-matrix 
using Korepin's method. We also reconsider the computation of 
the CS S-matrix in terms of particles with fractional charge. 

\end{abstract}

\maketitle

\section{Introduction}
Two-dimensional Conformal Field Theories (CFTs)\cite{cft}
and 
%(massive) 
Integrable Quantum Field Theories (IQFTs)\cite{Zam-Zam}
are usually treated in a completely different fashion despite 
the fact that they are deeply related \cite{zam.pert-cft,blz}.
A traditional starting point for the analysis of a CFT is the
Hilbert space for the system in finite geometry, organized in 
terms of representations of chiral algebras for left and right 
moving excitations. On the other hand, IQFTs are conveniently 
described in terms of asymptotic particle states and the
associated scattering data, pertaining to the system in infinite 
geometry. As a consequence of integrability the scattering is 
factorized, giving rise to a Factorized Scattering Theory (FST). 
The 2-particle S-matrix completely 
determines the on-shell dynamics as well as off-shell properties like 
correlation functions. In fact, once the exact particle spectrum and 
S-matrix are known,
one can use their spectral representations together with the knowledge of 
the exact matrix elements, obtained using the Form Factor
approach\cite{FF}, to compute correlation functions of local fields.
The factorized scattering approach is conceptually clear 
if the excitations in the field theory are massive, while, strictly
speaking, scattering of massless relativistic 
particles is not well defined in two dimensions. Nevertheless,  it
has been found that, if the massless scattering is suitably defined, 
the FST approach can be very fruitful also for massless IQFTs
\cite{zam.massless,zam-zam.massless,dms}. 

In the context of applications to condensed matter systems in one 
spatial dimension, formulations of CFT in terms of FST become natural,
both from a conceptual and from a computational point of view.
Such formulations involve the identification of a suitable set
of (massless) CFT quasiparticles with factorized scattering, and the 
study of their 2-particle $S$-matrix.
For a particular class of $c=1$ CFTs two interpretations in terms of
FST have been considered. The first one is closely related to 
the usual approach to massive FSTs \cite{zam.massless,zam-zam.massless};
the relevant S-matrix can be obtained as massless limit, $M\rightarrow 0$,
of the sine-Gordon (SG) S-matrix \cite{fsw}.
This particular description has been employed in the analysis of 
the edge-to-edge tunneling in fractional quantum Hall samples \cite{fls}.
The second approach is intrinsic to the CFT \cite{CFT.excl-stat,esp}
and leads to a description in terms of a gas of quasiparticles that satisfy 
fractional exclusion statistics \cite{haldane.excl-stat}. 
These quasiparticles have been identified with the excitations of the 
Calogero-Sutherland (CS) model in the continuum limit \cite{iso}, and 
their S-matrix was inferred on the basis of a set of Thermodynamic 
Bethe Ansatz equations \cite{iow,bw}.

The aim of this paper is to explore the relationship between these two 
FST descriptions of $c=1$ CFT. We will show that they emerge as different 
massless limits of the SG theory (see figure \ref{fig:figure}). 
In particular, the fractional statistics particles emerge as 
particle-hole excitations of the SG model in presence of a chemical 
potential, $A \to \infty$ \cite{abz}. We provide a direct calculation 
of their S-matrix using Korepin's method \cite{korepin}. We also 
reconsider the calculation of the CS S-matrix \cite{essler}
in terms of particles with fractional charge. 

Our presentation is organized as follows. In the next 
section we review the standard construction of interacting CFT 
quasiparticles via a massless limit of the SG theory. In Sec. III we 
introduce the fractional statistics quasiparticles in the CFT. 
Section IV is devoted to fractional excitations in the CS model. 
In Section V we show how the fractional statistics CFT quasiparticles 
can be obtained from the SG model in the presence of a chemical 
potential $A$, in the limit $A\to \infty$. We close with a brief 
discussion in Section VI.

\section{Massless limit of sine-Gordon: interacting quasiparticles}
\label{sec:sg}

A quasiparticle description of $c=1$ CFT can be obtained
by taking the massless limit of a SG model\cite{fsw}. In this 
section we recall a few basic facts about the SG theory and 
on the limit $M \to 0$. The SG action is given by
\begin{equation}
\label{sg}
{\cal S}_{SG} = \int \; d^2 x \left \{ \frac{1}{16\pi}(\partial_\nu
\varphi )^2 
-2 \mu \cos(\beta \varphi) \right \}\ ,
\end{equation}
where $\varphi(x)$ is a scalar field in 2D Euclidean space-time $x=(x^0,x^1)$.
This  model possesses a $U(1)$ symmetry, generated by the charge
\begin{equation}
Q=\int_{-\infty}^\infty \; j^0 \ dx=-\frac{\beta}{2\pi}
\int _{-\infty}^\infty \frac{\partial
\varphi}{\partial x}\ dx\ ,
\end{equation}
where 
$\label{j}
j^\mu=-\frac{\beta}{2\pi}\epsilon^{\mu \nu}\partial _\nu \varphi
$
is the Noether current.
% \footnote{It is normalized such that solitons and
%antisolitons have charges $\pm 1$.}.
From the perturbed CFT point of view, (\ref{sg}) can be considered 
as a Gaussian model
\begin{equation}
\label{sgauss}
{\cal S}_{Gauss}=\frac{1}{16\pi }\int \; d^2 x ~ (\partial_\nu
\varphi )^2
\end{equation}
perturbed by the relevant operator $\exp(\pm i \beta \varphi)$ of scale 
dimension $\Delta_\beta=\beta^2$. 

The SG model  is integrable and has
been studied in great detail over the last 25 years. 
%Some of the results obtained in the repulsive regime $\beta^2 >1/2$
%appear to depend on the regularization scheme 
%employed
%to deal with the UV divergences. Here we follow the method of \cite{jnw},
%which is very natural from a field-theory point of view.
The spectrum  depends on the value of the coupling constant
$\beta^2$ or, alternatively, on $
\xi=\frac{\beta^2}{1-\beta^2}$.
For $\beta^2<1$ the cosine term is relevant in the renormalization group
sense and dynamically generates a spectral gap $M$ in the
excitation spectrum. In the repulsive regime $1<\xi <\infty$ the
spectrum contains only charged particles of charge $Q=\pm 1$, which
are called solitons and antisolitons. 
The operators that generate these particles are\cite
{sol.creating.op,lz} 
\be
{\cal O}_{s,n}(x)=\re^{ i\frac{n}{4 \beta}\tilde \varphi(x)+
i \frac{s \beta}{n} \varphi(x)} \ ,
\label{o}
\ee
where the dual boson field, $\tilde \varphi$ is defined as
\be
\tilde \varphi(x)=\int_{-\infty}^x \p_y \varphi(x,y) \rd y \ .
\ee
These operators are in general non-local, and carry a spin $s$
and a topological charge $n$. For $s=1/2$, (\ref{o}) corresponds to 
the bosonization formula for 
fermions of the Massive Thirring model \cite{sol.creating.op}: $\Psi(x)=
{\cal O}_{\pm 1/2, 1},~ \Psi^*(x)={\cal O}_{\pm 1/2, -1}$.
 
In the attractive regime $0<\xi <1$ neutral soliton-antisoliton
bound states, $B_n,\; n=1,2,\ldots <1/\xi$, called breathers, 
are formed and the spectrum becomes more complicated. 
One usually distinguishes 
soliton, antisolitons and breathers by some internal indices 
$\epsilon=s,\bar s,B_n$. The two-particle S-matrix, 
$S_{\epsilon_1,\epsilon_2}^{\epsilon_1 ',\epsilon_2 '}$, have been 
known for some time \cite{smatrix:sg,Zam-Zam}. 
We report here the soliton-soliton
S-matrix, $S_{ss}=S_{ss}^{ss}$, 
that we need in the following
\be
S_{ss}(\theta)=-\exp [- i\delta_{ss} (\theta)] \ , 
\quad
\delta_{ss}(\theta)=\int_0^\infty
\frac{\sin(\theta t/\pi) \;
\sinh \left (
\frac{1-\xi}{2}t \right )}{t \; \cosh \left (\frac{t}{2} \right
) \; \sinh \left ( \frac{\xi t}{2} \right ) } dt \ .
\label{Sss}
\ee
Here the rapidity $\theta$ parametrizes the relativistic energy and 
momentum
($e(p)=\sqrt{p^2+M^2}$)
\begin{equation}
\label{ep} 
p=M\sinh\theta \ , \quad  e = M\cosh\theta.
\end{equation}
For large $|\theta|$, $\delta_{ss}(\theta)$ behaves like 
\be
\delta_{ss}(\theta) \simeq 
%\pm \delta_{ss}^\infty=
\pm 
\pi (\tilde p-1), ~~\theta\to \pm\infty
\label{delta.infty}
\ee
with $\tilde p=1/(2\beta^2)$.

An appropriate quasiparticle basis for the c=1 CFT 
can be obtained taking the 
massless limit of SG particles \cite{fsw,fls} described above. 
Formally this limit  is constructed shifting the rapidities 
$\theta \to \theta \pm \theta_0/2$ and taking the limits $\theta_0 \to 
+ \infty, 
\; M\to 0$ in such a  way that $m=M\exp(\theta_0/2)$ remains finite. 
In this way one obtains the massless dispersion relations 
$e=p=(m/2)\re^\theta$, for right (R) movers and  $e=-p=(m/2)\re^{-\theta}$
for left (L) movers (where R and L movers are defined as $p>0$ and $p<0$ 
branches of the massless dispersion relation $e=\pm p$). 
Taking the same limit on the S-matrix one finds that the 
quasiparticle spectrum remains  the same, i.e. it will have R and L 
solitons, antisolitons and breathers. While in the RR and LL sectors the 
S-matrix turns out to be the same as in the massive case, the RL (LR) 
scattering is trivial. This is obviously related to the conformal 
symmetry. The presence of non-trivial RL scattering would signal a flow 
between critical points \cite{zam.massless,zam-zam.massless}. 
The massless limit thus gives a FST of the CFT, with interacting 
quasiparticles with internal degrees of freedom and characterized by a 
non-diagonal S-matrix. 
The same result can be obtained by starting from the S-matrix axioms 
for unitarity, crossing and Yang-Baxter factorization directly for 
the massless particles \cite{zam-zam.massless}.

At criticality the boson field, $\varphi$, can be decomposed into its 
holomorphic and antiholomorphic parts as
\be
\varphi=\phi(z)+\bar \phi(\bar z)\ , \quad
\tilde \varphi=\phi(z)-\bar\phi(\bar z)
\ee
with $z=x+iy$ ($\bar z=x-i y$). The operators associated to massless solitons
(antisolitons) take the form
\be
\tilde{\cal O}_{s,n}=\re^{i \left ( \frac{s\beta}{n}+
\frac{n}{4\beta} \right )
\phi(z)+i\left (\frac{s\beta}{n}-\frac{n}{4 \beta} \right )
\bar\phi (\bar z)  }
\ee
and the chiral components 
\be
\tilde{\cal O}_{1/4 \beta^2,\pm 1}=\re^{\pm i \frac{1}{2\beta}  
\phi(z)} \ , \quad
\tilde{\cal O}_{-1/4\beta^2,\pm1}=\re^{\mp i \frac{1}{2\beta} 
\bar\phi (\bar z)}
\label{sco.massless}
\ee
correspond to the $U(1)$ conformal primary fields, $J$ and $\bar J$, 
introduced in the next section. In this case the boson is 
compactified with a compactification radius $R=1/(2\beta)$.

{\begin{figure}[ht]
\epsfxsize=8cm
\centerline{\epsfbox{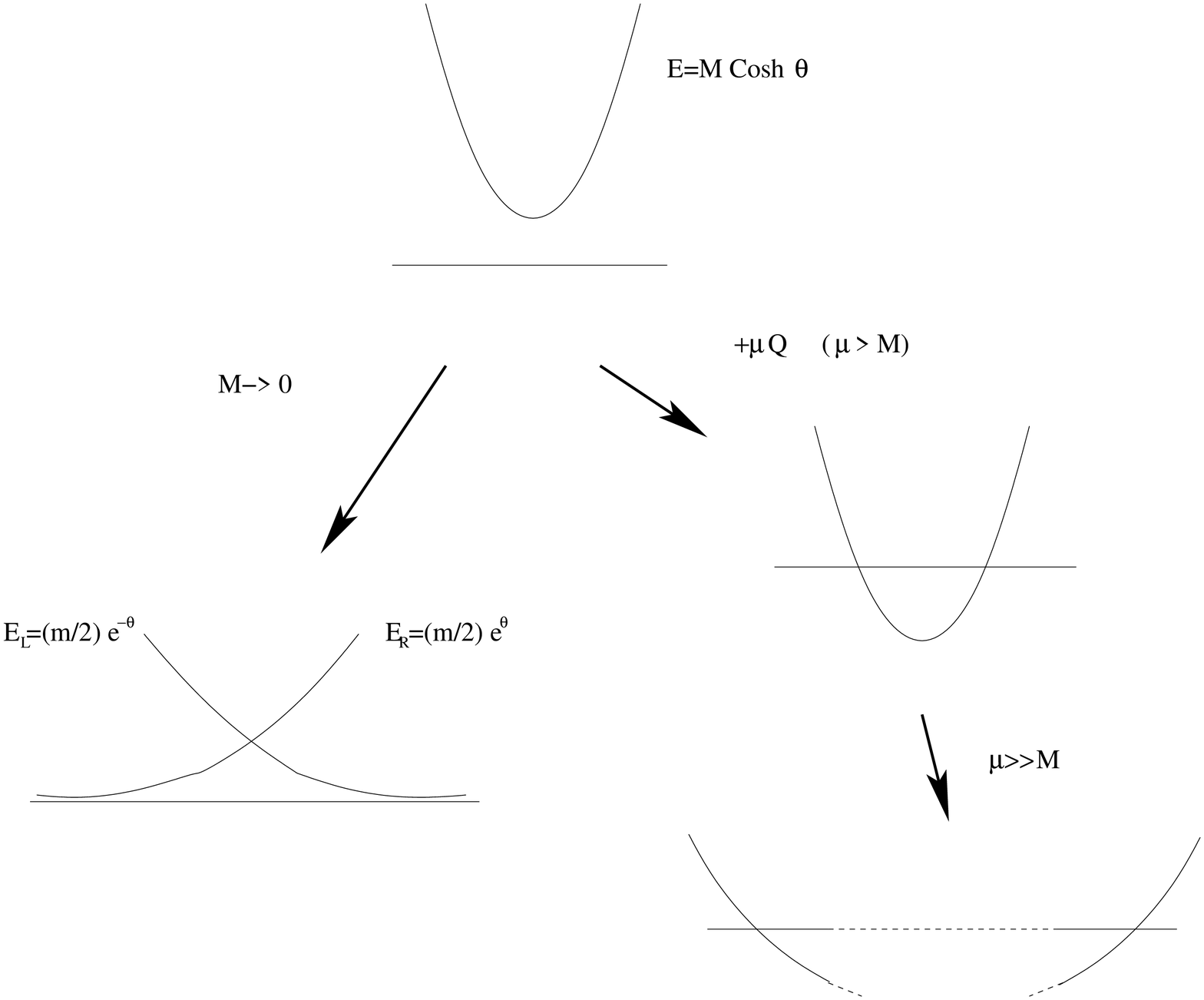}}
\caption{Schematic picture of the two different approaches to construct a 
quasiparticle representation of the $c=1$ CFT from the sine-Gordon model}
\label{fig:figure}
\end{figure} 
}

\section{CFT quasiparticles with fractional statistics}
\label{sec:FracStatCFT}

In this section we review the construction of a quasiparticle
basis for (chiral) $c=1$ CFT, which is very different from the
direct $M\to 0$ limit of the SG particle basis. This basis
is a special example of a `CFT quasiparticle basis' 
\cite{CFT.excl-stat}.
The idea behind the construction of the CFT quasiparticles is to
first classify the chiral primary fields, and then associate
quasiparticles to specific chiral primaries (usually those with
the smallest conformal dimensions). To construct an actual basis 
of the chiral Hilbert space one needs to impose specific rules on
the modes (momenta) of the quasiparticles that participate in a
multi-particle state; these rules then lead to a statement about
a form of exclusion statistics satisfied by the quasiparticles.

The prototype of a CFT quasiparticle basis is the so-called
spinon basis of the $c=1$ $SU(2)$ and $SU(N)$ invariant CFTs 
\cite{spinonCFT}; the more general construction was first outlined 
in \cite{CFT.excl-stat}. The particular
example we deal with here are the $c=1$ CFTs at compactification
radius $R^2=p/2$. In these theories, the chiral primary of smallest 
dimension carries a $U(1)$ charge $\pm {1 \over p}$, and one may 
contemplate quasiparticle bases built out of these fractionally 
charged quanta. This is particularly natural from the point of view 
of the fqH systems at filling fraction $\nu={1 \over p}$, which
have this particular CFT as effective edge theories \cite{esp}.

The most natural way to build a CFT quasiparticle basis at $c=1$,
$R^2=p/2$, involves quasiholes $\Psi_{qh}$, of charge $+{1 \over p}$, 
and particles $J$, of charge $-1$. They are described by the conformal
primary fields 
\begin{equation}
\label{jpsi}
J(z)=\re^{-i \sqrt{\frac{p}{2}}\phi(z)}=\sum_t J_{-t} z^{t-p/2} \qquad 
\Psi_{qh}
(z)=\re^{i {\frac{1}{\sqrt{2p}}}\phi(z)}=\sum_s \phi_{-s} z^{s-1/2p} \ 
\end{equation}
(note that the conventions used here are different from those in 
\cite{esp}).
The $J$ operators correspond to the massless limit of soliton creation
operators (\ref{sco.massless}) with the identification $p=\tilde p
=1/2\beta^2$.

The independent multi-particle states that generate the chiral Hilbert 
space were identified to be
\begin{eqnarray}
\label{eq:basis}
\lefteqn{
|m_M,\ldots m_1;n_N\ldots n_1>^Q \equiv}
\nonumber \\
&& J_{-(2M-1)p/2+Q-m_M} \ldots J_{-p/2+Q-m_1} \,  
\phi_{-(2N-1)/2p-Q/p-n_N} \ldots \phi_{-1/2p-Q/p-n1} |Q> \ ,
\end{eqnarray}
with
\be
m_M \geq \ldots \geq m_1 \geq 0 \ , \qquad 
n_N \geq \ldots \geq n_1 \geq 0 \qquad (n_1>0 \ 
{\rm if}\ Q<0),
\ee
where $|Q>$ ($Q=-(p-1),\ldots,-1,0$) is the lowest-energy state of charge 
$Q/p$. 

Using this basis, one can analyze the partition sum, and thereby the
thermodynamic properties, directly in terms of the quasiparticles
$J$ and $\phi$. One then finds that the thermodynamic equations
take the form of the so-called IOW equations \cite{iow} that describe 
the thermodynamics of a gas of particles satisfying fractional exclusion
statistics. For the general case with statistics matrix
${\bf g}=(g_{ij})$, the excitation energies, $\varepsilon_i$,
and distribution functions, $\bar{n}_i$, are determined  
by the following equations
\be
\left( \frac{\lambda_i-1}{\lambda_i} \right) 
\prod_j \lambda_j^{g_{ij}} = e^{\beta(\mu_i-\varepsilon_0)}\equiv z_i \ ,
\quad
\bar{n}_i(\varepsilon_i) = 
z_i \frac{\partial}{\partial z_i} \log \prod_j \lambda_j \ ,
\ee
with $\lambda_i=(1+\re^{-\varepsilon_i})$ being the one-particle grand 
canonical partition functions.
For the case of $c=1$ one finds the following statistical parameters:
%These thermodynamic equations  are precisely
%of the IOW type, with statistics parameters
$g_{\phi\phi}=1/p$, $g_{JJ}=p$ and $g_{\phi J}=g_{J\phi}=0$.

Summarizing, we see that the $c=1$, $R^2=p/2$ chiral CFT is described
by quasiparticles with charge/statistics parameters
$(Q={1 \over p},g={1 \over p})$ and $(Q={-1},g=p)$ and with
no mutual statistics between the two.

It is well-known \cite{iow,bw} that the IOW equations for statistics
matrix $g_{ij}$ agree with the TBA equations for a `purely
statistical' 2-particle $S$-matrix given by
\be
S_{ij}(\theta) 
=- \exp \left[ \pi i(\delta_{ij}-g_{ij}) {\rm sign}(\theta) \right] \ .
\ee
where ${\rm sign}(x)=|x|/x$ is the sign function (${\rm sign}(0)=0$).
Combining this with the above, we tentatively identify the above
CFT quasiparticle basis with a particle basis in the sense of
FST. In the remainder of this paper, we substantiate this claim
by establishing the relation between the CFT and this particular 
FST from two alternative points of view, which are the 
Calogero-Sutherland (CS) model and an alternative ($A\to\infty$)
massless limit of the SG model.

\section{S-matrix for fractional excitations in the CS
model}
\label{sec:CS}

The CFT quasi-particles described in the previous section are particularly
natural if the CFT is viewed as the continuum limit of a so-called 
Calogero-Sutherland (CS) model for particles with inverse-square interaction,
as the CFT quasiparticles can be identified with the fundamental 
excitations of the CS model \cite{iso}. In this section we evaluate
the S-matrix of the CS quasi-particles using a method developed by
Korepin. A related calculation was done in Ref.\cite{essler};
here we use a different scheme based on excitations with fractional 
charge.

The CS model \cite{sutherland1,sutherland2,calogero}
describes fermionic particles on a line 
whose interaction, for $N$ particles, is given by the following Hamiltonian
\be
H=-\sum_{j=1}^{N} \frac{\p^2}{\p x_j^2}
+\sum_{j<k} \frac{2p(p-1)}{(x_k-x_j)^2} \ .
\ee
In the low energy sector it is described by $c=1$ CFT with compactification 
radius that, using the conventions of Sec. \ref{sec:FracStatCFT}, is
$R^2=p/2$ \cite{iso}. 
The model was solved with the Bethe Ansatz (BA) in Ref.\cite{sutherland2}.
Imposing periodic boundary conditions on the wave function, one
obtains the following quantization (BA) equations\cite{sutherland2} 
\be
L \lambda_i=2 \pi I_i + \sum_{j\neq i}^{N}
\theta (\lambda_i-\lambda_j),~~~i=1,\ldots,N
\label{ba}
\ee
where 
\be
\theta(\lambda)=(p-1) \pi \, {\rm sign}(\lambda).
\ee
%with ${\rm sign}(\lambda)=\lambda/|\lambda|$ being the sign function.
The set of integers or half-odd integers $I_i$ is in one-to-one
correspondence with a set of spectral parameters that specify an
eigenstate of the Hamiltonian. The total energy 
is parametrized as:
$E=\sum_i\lambda_i^2$. In the presence of chemical potential, $\mu$,
the ground state is obtained by filling the Fermi sea with rapidities
$|\lambda|<\sqrt\mu\equiv \lambda_F$, 
at distances $\lambda_i-\lambda_j > \frac{2 \pi (p-1)}
{L}$ (this is the first signal of a generalized exclusion statistics). 
The Fermi momentum $\lambda_F$ can be written in terms of the total 
number of $\lambda$s in the condensate, $N_{GS}$, as: 
$\lambda_F=\pi p (N_{GS}-1)/L$.
%Then the total number of $\lambda$s will be given by: 
%$N_{GS}=\left [ \lambda_F L/\pi p \right ]$ (where $\left[ a \right ]$ is 
%the integer part).
The integers (half-odd integers) $I_i$ vary in the interval
\be
%\label{imax}
|I_j|<I_{max}^{GS}
\ee
with
\be
I^{GS}_{max}=\frac{1}{4\pi}|\lambda_F L - \sum_{k}^{N_{GS}}
\theta (+\infty)|=\frac{1}{2}(N_{GS}-1) \ .
\label{imax}
\ee
There are different types of possible excitations.

\underline{I: Quasihole excitations}.
%This excitations
The lowest energy hole-type excitation corresponds to
occupation numbers $N=N_{GS}-1$. If we keep the Fermi momentum, $\lambda_F$,
fixed to order one, the range of
the integers $I_i$ changes according to (\ref{imax}) 
\be
I_{max}=I_{max}^{GS}+\frac{p-1}{2} \ .
\ee
Then one has $p-1$ additional vacancies and one less $\lambda$, i.e. the
excitations are characterized by $p$ parameters. We can say that this
excitation
corresponds to creating $p$ quasiholes, $\lambda_h^1,\ldots,\lambda_h^p$, 
in the ground state, each having charge $1/p$. 
We use the term quasihole to distinguish these excitations 
from the charge-$1$ hole excitations that we will discuss in the following.
The $p$-quasihole excitations are similar to the $S=1$ 2-spinon excitations 
in the $S=1/2$ Heisenberg \cite{heisenberg} or Haldane-Shastry models
\cite{essler}.

The BA equations relative to this type of excitation are the following
\be
L \tilde\lambda_i=2 \pi \tilde I_i + \sum_{j\neq i}^{N_{GS}+(p-1)}
\theta (\tilde \lambda_i-\tilde \lambda_j)-\sum_{a=1}^p \theta (\tilde
\lambda_i -\lambda^h_a)
\label{ba.hole}
\ee
where $\tilde \lambda$ indicates that the
spectral parameters have slightly changed with respect to the ground state
distribution ($\tilde\lambda_i-\lambda_i \sim O(1/L)$).
We follow the conventions of
\cite{ek}, take $\tilde I_i - I_i=(p-1)/2$ for $i=1,\ldots,N_{GS}$,
and place the additional vacancies,
$\lambda_1^{min},\ldots,\lambda_{p-1}^{min}$,
close to the left Fermi point. In the thermodynamic limit $\lambda_i^{min}
\to - \infty$.
As usual, we characterize these excitations via the shift function
\cite{korepin}
\be
F_{hh}(\lambda_i)=\frac{\tilde \lambda_i -\lambda_i}
{\lambda_{i+1}-\lambda_i} \ .
\ee
The equation defining $F_{hh}(\lambda)$ 
can be obtained subtracting (\ref{ba.hole}) from the
ground state distribution and using the fact that
$\tilde\lambda_i-\lambda_i \sim O(1/L)$,
\be
F_{hh}(\lambda_i)+\frac{1}{2\pi} \sum_{j\neq i}^{N_{GS}}
K(\lambda_i-\lambda_j)F_{hh}(\lambda_j) 
=\frac{p-1}{2}+\frac{1}{2\pi} \sum_{k=1}^{p-1}
\theta (\lambda_i-\lambda_k^{min}) -
\frac{1}{2\pi} \sum_{a=1}^{p} \theta(\lambda_i-\lambda^h_a) \ ,
\ee
where $K(\lambda)=\theta'(\lambda)=2\pi(p-1)\delta(\lambda)$. 
In the thermodynamic limit
it becomes
\be
\label{Fhh0}
F_{hh}(\lambda)+\frac{1}{2\pi} \int_{-\lambda_F}^{\lambda_F} \rd \mu
K(\lambda-\mu)F_{hh}(\mu)=
\frac{p-1}{2}+\frac{1}{2\pi} (p-1)\theta (+\infty) -
\frac{1}{2\pi} \sum_{a=1}^{p} \theta(\lambda-\lambda^h_a) \ .
\ee
Using the explicit form of $K(\lambda)$ and $\theta(\lambda)$
we find
\be
F_{hh}(\lambda_h^1)=\frac{p-1}{2p}, ~~ \lambda_h^1> \lambda_h^b 
~(b=2,\ldots,p) \ .
\label{Fhh}
\ee

\underline{II: Particle excitations}.
Particle excitations correspond to $N=N_{GS} +1$,
where the additional particle has rapidity
$|\lambda_p|>\lambda_F$. The BA equations have the form
\bea
&&L \tilde \lambda_i=2 \pi \tilde I_i + \sum_{j\neq i}^{N_{GS}}
\theta (\tilde \lambda_i-\tilde \lambda_j) +\theta(\tilde \lambda_i 
-\lambda_p) \ ,
\\
&&L \lambda_p=2 \pi \tilde I_{N_{GS}+1} + \sum_{j}^{N_{GS}}
\theta (\lambda_p-\tilde \lambda_j) \ .
\eea
In the thermodynamic limit the integral equations for $F$ have the form
\be
F_{pp}(\lambda)+\frac{1}{2\pi} \int_{-\lambda_F}^{\lambda_F}\rd \mu
K(\lambda-\mu)F_{pp}(\mu)
=\theta(\lambda-\lambda_p)
\ee
with the explicit solution
\be
F_{pp}(\lambda_p)=\frac{(p-1)}{2} \ .
\label{Fpp}
\ee

From these two types of excitations it is possible to construct neutral
excitations corresponding to $1$ particle and $p$ quasiholes.
The shift function associated to these types of excitations is
defined by the equation
\bea
\lefteqn{
F_{ph}(\lambda)+\frac{1}{2\pi}  \int_{-\lambda_F}^{\lambda_F}\rd \mu
K(\lambda-\mu)F_{ph}(\mu)=}
\nonumber\\[2mm] &&
\frac{p-1}{2}+\frac{1}{2\pi}\theta(\lambda-\lambda_p)+  
\frac{1}{2\pi} (p-1)\theta (+\infty) -
\frac{1}{2\pi} \sum_{a=1}^{p} \theta(\lambda-\lambda^h_a) 
\eea
from which
\be
F_{ph}(\lambda_p)=0,~~{\rm mod}2\pi \ .
\label{Fph}
\ee
We choose these types of excitations to describe the Hilbert space 
and construct the FST. Below we explicitly calculate their S-matrix.

%\underline{III: Particle and hole excitations with integer charge}.
In Ref.~\cite{essler} the physical excitations were constructed using a 
different scheme that corresponds to allowing the Fermi momentum in 
(\ref{imax}) 
to vary of order $O(1/L)$ when one creates a hole excitation. Within this
scheme, changing the occupation numbers by one 
($N_{GS}\to N_{GS}-1$) produces a
shift of $I_{max}$ by one. In this way  all the excitations have 
integer charge. 
%Neutral excitations can be obtained also as particle-hole excitations. 
%These type of excitation can be constructed moving
%a rapidity $\lambda_h < \lambda_F$ to $\lambda_p>\lambda_F$. By construction
%these excitations have the same quantum number as in the ground state
%and are characterized by the following BA equations
%\be
%L \tilde\lambda_i=2 \pi \tilde I_i + \sum_{j\neq i}^{N_{GS}}
%\theta (\tilde \lambda_i-\tilde \lambda_j)
%+\theta(\tilde \lambda_i - \lambda_p) -\theta(\tilde \lambda_i - \lambda_h)
%\ee
%We note that this excitations are present only in presence of a chemical
%potential and provide a
%different scheme to describe the
%excitations and construct the Hilbert space,
%where both particle and holes have charge 1 \cite{essler}.
%Within this scheme  a single particle or hole can be obtained from a 
%particle-hole excitation placing the
%other one at the Fermi level.  
On the level of the BA the two descriptions seem to be 
complementary. For reasons
that will become clear later we prefer to use as basic
excitations I and II, i.e. particles with integer charge but quasiholes 
with a fractional charge.
This approach is closer to the philosophy of Ref. \cite{ha,lps}.

\underline{S-matrix for CS model}.
We now apply Korepin's method \cite{korepin}
to compute the S-matrix relative to the excitations I and II described
above. The S-matrix is just a phase
\be
S_{ab}= \exp (- i \delta_{ab}),
\ee
where the two-particle scattering phase is equal to the phase, $\varphi_{ab}$,
obtained by
moving the particle $a$ through the system in presence of particle $b$
minus the phase shift, $\varphi_{a}$, obtained through the same process but in
absence of particle $b$
\be
\delta_{ab}=\varphi_{ab}-\varphi_a.
\ee 
Both phases can be obtained using the BA equations for the ground state and
excitations.
%and the excitations considered above. It turns out \cite{korepin} 
In general, it turns out \cite{korepin} that  
$\delta_{ab}$
is related to the shift function via
\be
\delta_{ab}(\lambda_i,\lambda_j) |_{\lambda_i>\lambda_j}=
2 \pi F_{ab}(\lambda_i) \ .
\label{deltaF}
\ee
Unfortunately, this approach cannot be applied directly to the 
quasihole excitations introduced in the previous section. 
%In particular 
%some care must be taken when considering quasihole excitations.
In fact, within the scheme we consider here,
it is not possible to create one and two-quasihole excitations
and thus it is not possible to evaluate directly the
one and two-particle phase shift. Nevertheless it is possible to
evaluate the total  phase shift associated to  an excitation I
(consisting of $p$ quasiholes) as a whole.  
This corresponds to the phase shift acquired by the fastest particle going 
across the other $p-1$ quasiholes. 
Using factorization, together with the fact that the resulting
S-matrix is momentum independent, 
%the
%p-particle S-matrix can be written as
%\bea
%S_{p,hh}(\lambda_1,\ldots,\lambda_p)=\prod_{i<j}
%S_{2,hh}(\lambda_i,\lambda_j)\simeq \re^{\sum_{i<j} \delta_{hh}
%(\lambda_i,\lambda_j)}\simeq 
%\re^{(p-1) \delta_{hh}}
%\eea
%Then  we have
one can see that the quasihole-quasihole
phase shift, $\varphi_{hh}$,  is given by
\be
\varphi_{hh}(\lambda_h^1,\lambda_h^2)|_{\lambda_h^1>\lambda_h^2}
= 2 \pi F_{hh}(\lambda_h^1)/(p-1)= \pi/p ~~~
(\lambda_h^1>\lambda_h^b, b=2,\ldots,p).
\ee
In order to compute $\delta_{hh}$ we should now subtract the one particle
phase shift, $\varphi_h$.  Having no direct access to $\varphi_h$,
we subtract a reference phase, $\delta_0(p)$,  
\be
\delta_{hh}(\lambda_h^1,\lambda_h^2)|_{\lambda_h^1>\lambda_h^2}
= 2 \pi F_{hh}(\lambda_h^1)/(p-1)-\delta_0(p)= \pi/p-\delta_0(p) ~~~
(\lambda_h^1>\lambda_h^b, b=2,\ldots,p) \ .
\label{delta.hh}
\ee
Based on the observation that individual quasi-holes are local with 
respect to the CS ground state, we anticipate that this reference phase 
will be an integer multiple of $\pi$. Below we shall fix its value by 
an independent argument.

Followign a similar reasoning we find the particle-quasihole phase shift,  
\be
\delta_{ph}(\lambda_p,\lambda_h^1)= 2 \pi 
F_{ph}(\lambda_p)/p = 0 ~{\rm mod}2\pi \ .
%\;  .
\label{delta.ph}
\ee
The particle-particle S-matrix can be computed in a more standard 
way\cite{essler} since  
one and two-particle excitations can be constructed explicitly. As a
consequence, the two-particle S-matrix will be completely determined. 
One finds
\be
\delta_{pp}(\lambda_p^1,\lambda_p^2)|_{\lambda_p^1> \lambda_p^2}
= 2 \pi F_{pp}(\lambda_p^1)
= (p-1)\pi.
\label{delta.pp}
\ee
The identification of (\ref{delta.hh}, \ref{delta.ph}, \ref{delta.pp}) 
as phase shifts clearly requires
$\lambda_1>\lambda_2$. Analytic continuation of these results to 
the sector $\lambda_1<\lambda_2$ can be done using the unitarity condition 
of the S-matrix: $S_{ab}(\lambda)S_{ab}(-\lambda)=1$. From which it follows
\be
S_{ab}(\lambda)=- \exp \left[ - i \delta_{ab}~{\rm sign}( \lambda) \right ],
~~~a,b=p,h
\ee
where we have chosen a ``fermionic'' 
normalization: $S_{ab}(0)=-1$\cite{zamTBA}.
In order to fix the reference phase, $\delta_0(p)$, in (\ref{delta.hh})
we use the known duality of the CS Hamiltonian that, under $p\to1/p$, maps
particles into holes and vice-versa\cite{duality}.  
This implies $\delta_0(p)=\pi$, in agreement with our expectation.
Our full result for the 2-particle S-matrix is in agreement with what
was found using different methods \cite{esp} as reported in 
Sec.~\ref{sec:FracStatCFT}.

We note that the quasihole-quasihole scattering phase has a dependence
typical of particles with
fractional statistics, this is the reason for the choice of the
scheme above. Our results are consistent, we have particles with fractional
charge that satisfy fractional exclusion statistics.

\section{Fractional CFT quasiparticles from sine-Gordon}

Let us now consider the  SG model in presence of a chemical potential
coupled to the conserved charge. The Hamiltonian is shifted as
\be
\label{sg.A}
{\cal H}(A)={\cal H}_{SG}- A Q
\ee
The potential $A$ works as infrared cut-off at scales of the order $A$, and 
therefore for $A\gg \mu^{(p+1)/2}$ the theory is driven to the UV fixed
point where it is described by the $c=1$ CFT (\ref{sgauss}).
In presence of the chemical potential every soliton (antisoliton) 
acquires an additional energy $A(-A)$, while the breathers spectrum in
not affected.
For $A > M $ the ground state is a soliton condensate and theory has 
massless excitations across the Fermi sea. The other excitations have a 
gap, so we do not need to consider them for what follows.
Using Korepin's method \cite{korepin}, we shall construct the S-matrix 
for excitations over the soliton condensate in the limit $A\to \infty$, 
and show that they are free particles satisfying fractional statistics. 
These quasiparticles then provide a FST description of the $c=1$ fixed 
point. We will find that the S-matrix is the same as the one for
the CS model constructed in the previous section. 

We can repeat the analysis of the excitations and S-matrix
done for the CS model in Sec.\ref{sec:CS} for the SG model in presence of
the chemical potential $A>M$ (\ref{sg.A}). Putting $N$ solitons on thespace
line of length $L$ and imposing periodic boundary conditions one obtains 
the following quantization equations\cite{zamTBA}
\be
L p(\theta_i)=2 \pi I_i + \sum_{j\neq i}^{N}
\delta_{ss} (\theta_i-\theta_j), ~~~i=1,\ldots,N
\label{ba.sg}
\ee
where $\delta_{ss}(\theta)$ and $p(\theta)$ are defined in (\ref{Sss}) 
and (\ref{ep}) respectively. The presence of the chemical potential
induces also in this case a particle condensate and the  
ground state is obtained filling the rapidities symmetrically around zero. 
In the thermodynamic limit
the ground state energy in presence of the chemical potential is given by
\be
{\cal E}(A)-{\cal E}(0)=\frac{M}{2\pi}\int_{-B}^{B} \rd \theta
\cosh \theta\; \epsilon(\theta) \ ,
\label{gs}
\ee
where $\epsilon(\theta)$ is a non-positive function 
defined by the following equation
\be
\epsilon(\theta)+
\int_{-B}^{B} \rd \theta'
 K_{ss}(\theta-\theta') \epsilon(\theta')=M\cosh\theta-A, 
~\epsilon(\pm B)=0
\label{epsilon}
\ee
with 
\be
 K_{ss}(\theta)=\frac{1}{2\pi}
\frac{\rd \delta_{ss}(\theta)}{\rd \theta}.
\ee
In  momentum space the kernel 
$\tilde K_{ss}(\theta)=\delta(\theta)+K_{ss}(\theta)$ 
has a quite simple form
\be
\tilde K_{ss}(\omega)=\int_{-\infty}^{\infty} \rd \theta \re^{i\omega \theta}
\tilde K_{ss}(\theta)=\frac{\sinh\frac{\pi(1+\xi)\omega}{2}}
{2\cosh\frac{\pi \omega}{2} \sinh\frac{\pi \xi\omega}{2}}
\ee
and can be factorized as 
\be
\tilde K_{ss}(\omega)=\frac{1}{K_+(\omega)K_-(\omega)}
\ee
with 
\be
K_+(\omega)=K_-(-\omega)=\sqrt{\frac{2\pi(\xi+1)}{\xi}}\re^{i\omega \Delta}
\frac{\Gamma\left(i\frac{\xi+1}{2}\omega\right )}{
\Gamma\left(i\frac{i\xi\omega}{2}\right ) \Gamma\left ( \frac{1}{2}+
\frac{i\omega}{2} \right )}
\label{kpm}
\ee
analytic in the upper ($K_+$) and lower ($K_-$) half plane. In (\ref{kpm})
$\Delta=\frac{\xi}{2}\log \xi -\frac{\xi+1}{2}\log (\xi+1)$ so that 
$K_+(\omega)=1+O(1/\omega)$.
The limit $A\to \infty$ corresponds to $B\to \infty$ and 
Eqs.(\ref{gs},\ref{epsilon}) become \cite{abz}
\bea
&&{\cal E}(A)-{\cal E}(0)=\frac{M}{4\pi}\int_{-B}^{B} \rd \theta
\re^\theta\; \epsilon(\theta)\\[2mm]
&&
\int_{-\infty}^{B} \rd \theta'
\tilde K_{ss}(\theta-\theta') \epsilon(\theta')=\frac{M}{2}\re^\theta -A,
\eea
clearly describing a massless system.

The excitations can be studied using (\ref{ba.sg}).
They turn out to be the similar to those discussed in Sec. \ref{sec:CS} 
for the CS model. Also in this case we find quasihole excitations with 
fractional
charge,  the reason being that $\delta_{ss}(+\infty)\neq 1$, 
%(cf. Eq. (\ref{delta.infty})) 
and removing a
$\theta$ produces a shift of $I_{max}$ 
according to (\ref{imax}).
From Eq. (\ref{delta.infty}) we see that this shift produces $\tilde p -1$ 
additional vacancies. Then again an excitation will be characterized 
by $\tilde p$ parameters and can be interpreted as consisting of $\tilde p$ 
quasiholes of charge $1/\tilde p$ in the ground state.
Below we show that, in the limit $A \to \infty$,
the S-matrix relative to these excitations
is again given by (\ref{delta.hh}), (\ref{delta.ph}) and (\ref{delta.pp}),
where $\tilde p$ replaces the CS parameter $p$. With this identification 
the CS model and the UV limit of the SG model give rise to the
same quasiparticle S-matrix. For any finite $A$ there are
corrections to the CS S-matrix that depend on the rapidities.

Let us first consider quasihole excitations. They will be characterized by the
following BA equations
\be
L p(\tilde\theta_i)=2 \pi \tilde I_i + \sum_{j\neq i}^{N_{GS}+(\tilde p-1)}
\delta_{ss} (\tilde \theta_i-\tilde \theta_j)-\sum_{a=1}^p \delta_{ss} 
(\tilde\theta_i -\theta_h^a)
\label{sg.ba.hole}
\ee
and can be studied again introducing  the
shift-function, $\tilde F_{hh}(\theta_i)=
(\tilde \theta_i-\theta_i)/(\theta_{i+1}-\theta_i)$, satisfying the following
equation (in the thermodynamic limit) 
\be
\tilde F_{hh}(\theta)+\frac{1}{2\pi} \int_{-B}^{B} \rd \theta '
K_{ss}(\theta-\theta ')\tilde F_{hh}(\theta ')=
\frac{\tilde p-1}{2}+\frac{1}{2\pi} (\tilde p-1)\delta_{ss} (+\infty) -
\frac{1}{2\pi} \sum_{a=1}^{\tilde p} \delta_{ss}(\theta-\theta_h^a) \ . 
\label{sgFhh}
\ee
%Following the reasoning of Sec. 
%\ref{sec:CS} one can show  
%that the phase shift from quasihole-quasihole 
%scattering, $\tilde \delta_{hh}$,
%is again related to the the shift function by:
%$\tilde \delta_{hh}(\theta_h^1,\theta_h^2)|_{\theta_h^1>\theta_h^2}=
%2 \pi \tilde F_{hh}(\theta_h^1)/(\tilde p-1)$ (with  
%$\theta_h^1 > \theta_h^b,~~~b=2,\ldots,\tilde p$) .
In the limit $B\to \infty$, Eq. (\ref{sgFhh}) can be 
solved with the Wiener-Hopf
method (see for instance Appendix B of Ref.\cite{abz}). The procedure is 
quite standard and we report here only the essential steps.
Shifting $\theta_h^1\to \theta_h^1+B$
and introducing $f_{hh}(\theta_h^1)=F_{hh}(\theta_h^1+B)$ 
one can rewrite 
(\ref{sgFhh}) as
\be
\label{fhh00}
f_{hh}(\theta_h^1)+\frac{1}{2\pi} \int_{-2B}^{0} \rd \theta '
K_{ss}(\theta_h^1-\theta ')f_{hh}(\theta ')=
\frac{\tilde p-1}{2}+\frac{1}{2\pi} (\tilde p-1)\delta_{ss} (+\infty)
-\frac{1}{2\pi} \sum_{a=1}^{\tilde p-1} \delta_{ss}(\theta_h^1-\theta^h_a+B). 
\ee
As a consequence of the shift, quasihole excitations now correspond to
$\theta_1^h<0$. 
For $B\to \infty$ Eq.~(\ref{fhh00}) takes the form
\be
f_{hh}(\theta_h^1)+\frac{1}{2\pi} \int_{-\infty}^{0} \rd \theta '
K_{ss}(\theta_h^1-\theta ')f_{hh}(\theta ')=g_\infty(\tilde p) \ ,
\label{fhh}
\ee
where on the RHS we have approximated $\delta_{ss}(\theta)$ 
with its asymptotics
\be
g_\infty(\tilde p)= \frac{\tilde p-1}{2}+\frac{1}{2\pi} (\tilde p-1)
\delta_{ss} (+\infty)
-\frac{1}{2\pi} \sum_{a=1}^{\tilde p-1} \delta_{ss}^\infty=
\frac{\tilde p-1}{2} \ . 
\ee
This is the same driving term as for the CS model (\ref{Fhh0}).
We can solve this equation for any $\theta$ although we only need it 
for $\theta<0$.
We  rewrite 
$f(\theta)$  as   
$f(\theta)=f^+(\theta)+f^-(\theta)$, where $f^+(\theta)=f(\theta)$ 
for $\theta>0$ and zero otherwise, and 
$f^-(\theta)=f(\theta)$ for $\theta<0$. 
Fourier transforming, Eq. (\ref{fhh}) becomes
\be
f^+(\omega)+\tilde K_{ss} (\omega) f^-(\omega)
=g_\infty(\tilde p) \delta(\omega)
\ee
and can be solved with the WH method, to give 
\be
f^+(\omega)=g_\infty(\tilde p)\frac{K_+(0)}{K_+(\omega)}\frac{1}{\omega+i0}
\ , \quad
f^-(\omega)=g_\infty(\tilde p) K_+(0)\frac{K_-(\omega)}{\omega-i0} \ .
\label{fminus}
\ee
We can now  obtain $F_{hh}(\theta_h^1)$ by Fourier transforming Eq. 
(\ref{fminus})
\be
F_{hh}(\theta_h^1)=\int \rd \omega \re^{i \omega (\theta_h^1+B)} 
f^-(\omega)
\sim g_\infty(\tilde p) K_+(0) K_-(0)=\frac{g_\infty(\tilde p)}
{\tilde K_{ss}(0)}=
\frac{\tilde p-1}{2 \tilde p}, 
\ee
where we have omitted terms of order $O(\exp(-B))$.
%From here it follows that in the limit $A\to \infty$
%\be
%\tilde \delta_{hh}=\pi /p
%\ee
This  is the same as for the CS model with the identification 
$\tilde p=p$, as previously anticipated.
Following the same steps as in Sec.\ref{sec:CS}, we obtain also $\tilde 
\delta_{ph}= 0$ and $\tilde 
\delta_{pp}=(\tilde p -1)\pi$. 
Although it is not possible to show it on the basis of the BA, it is quite 
natural to argue that these excitations are generated by the operators
(\ref{jpsi}).

\section{Discussion}
In this paper we showed how fractional statistics quasiparticles
in specific $c=1$ CFTs can be obtained from an associated sine-Gordon 
model. Introducing a chemical potential, $A$, and driving the system to 
the UV fixed point by taking $A \to \infty$, we constructed massless 
excitations with fractional charge and computed their S-matrix. These 
excitations correspond to the excitations of the Calogero-Sutherland 
model associated to the same CFT. Their S-matrix is momentum 
independent, giving rise to the notion of a free gas of particles with 
generalized statistics. This formulation of the $c=1$ CFT can be
contrasted with the formulation obtained via a massless limit,
$M \to 0$, of the same sine-Gordon theory. Our result thus sheds 
some light on the relation between different Factorized Scattering
Theories (FST) associated to a $c=1$ CFT. It will be worthwhile
to explore similar relations for FST formulations of more general
(rational) CFTs.

\section{Acknowledgments}
We would like to thank F. Essler for important
discussions and clarifications on his work on the Calogero-Sutherland 
model. We also thank J-S. Caux and S. Peysson for useful discussions.
D.C. is supported by the European Community under Marie Curie Fellowship
grant HPMF-CT-2002-01591. K.S. is supported in part by the foundations 
FOM and NWO of the Netherlands.

\end{document}